\documentclass[runningheads]{llncs}
\usepackage{graphicx, amssymb, amsmath}
\begin{document}
\title{Variational Autoencoders for Precoding Matrices with High Spectral Efficiency}
\titlerunning{Variational Autoencoders for Precoding Matrices}
%
\author{Evgeny~Bobrov\inst{1, 2}\orcidID{0000-0002-2584-6649 } \\
Alexander~Markov\inst{5} \and Sviatoslav~Panchenko\inst{5} \and Dmitry~Vetrov\inst{3, 4}}
\authorrunning{E. Bobrov et al.}

%
\institute{M. V. Lomonosov Moscow State University, Russia \\
\email{eugenbobrov@ya.ru} \and
Moscow Research Center, Huawei Technologies, Russia \and
National Research University Higher School of Economics, Russia \and
Artificial Intelligence Research Institute, Russia \\
\email{vetrovd@yandex.ru} \and
Moscow Institute of Physics and Technology, Russia \\
\email{almarkv@yandex.ru, panchenko.sk@phystech.edu}}
\maketitle              
\begin{abstract}

Neural networks are used for channel decoding, channel detection, channel evaluation, and resource management in multi-input and multi-output (MIMO) wireless communication systems. In this paper, we consider the problem of finding precoding matrices with high spectral efficiency (SE) using variational autoencoder (VAE). We propose a computationally efficient algorithm for sampling precoding matrices with minimal loss of quality compared to the optimal precoding. In addition to VAE, we use the conditional variational autoencoder (CVAE) to build a unified generative model. Both of these methods are able to reconstruct the distribution of precoding matrices of high SE by sampling latent variables. This distribution obtained using VAE and CVAE methods is described in the literature for the first time.


\end{abstract}
\section{Introduction}
Massive multiple-input multiple-output (MIMO) is one of the core technologies of the fifth-generation (5G) wireless systems as it promises significant improvements in spectral efficiency (SE)~\cite{SE}, compared to the traditional, small-scale MIMO. Massive MIMO scales up the number of antennas in a conventional system by orders of magnitude~\cite{rusek2012scaling}. In multi-user (MU) massive MIMO, the infrastructure base stations (BSs) are equipped with hundreds of antenna elements that simultaneously serve tens of user equipments (UEs) in the same frequency band. 

A key advantage of the massive number of antennas is that the SE quality is significantly improved by employing simple linear signal processing techniques ~\cite{bjornson2016massive},~\cite{larsson2014massive},~\cite{abdallah2018mmse},~\cite{albreem2019massive}. In multiple-input multiple-output (MIMO) systems with a large number of antennas, precoding is an important part of downlink signal processing, since this procedure can focus the transmission signal energy on smaller areas and allows for greater spectral efficiency with less transmitted power~\cite{EE},~\cite{5G},~\cite{MCS}. 

Various linear precodings allow to direct the maximum amount of energy to the user as Maximum Ratio Transmission (MRT) or completely get rid of the inter-user interference as Zero-Forcing (ZF)~\cite{ZF_MRT}. There are also different non-linear precoding techniques such as Dirty Paper Coding (DPC), Vector Perturbation (VP), and L-BFGS SE Optimization, which achieve better SE quality than linear methods but have higher implementation complexity~\cite{DPC}. The authors of L-BFGS SE Optimization reduce the task of finding precoding matrices with high spectral efficiency to an unconstrained optimization problem using a special differential projection method and solve it by the Quasi-Newton iterative procedure to achieve gains in capacity~\cite{LBFGS}. We take this work as a basis and accelerate the solution of the iterative procedure using machine learning methods.

The application of machine learning in the field of wireless communication has been widely studied~\cite{zhao2021variational},~\cite{xia20},~\cite{bobrov2021machine},~\cite{zhao2020deep} and deep neural network technology has recently become a promising tool in the field of wireless physical layer technology~\cite{ye2017power}.  In~\cite{ye2019deep}, deep reinforcement learning is applied to a new decentralized resource allocation mechanism for communication between vehicles. In~\cite{balatsoukas2019neural}, deep learning tools are used to automatically configure 1-bit precoders in complex MU precoding algorithms.  The~\cite{van2020deep} offers a deep energy autoencoder for incoherent MU-MIMO systems with multiple carriers, achieving higher reliability and flexibility.~\cite{dong2020framework} has developed a new collaborative hybrid processing platform based on deep learning for end-to-end optimization. A beam selection method based on a deep neural network is proposed by~\cite{rezaie2020location}. In~\cite{huang2020deep}, the choice of a delayed repeater based on deep learning for secure cognitive relay networks with buffer management is investigated. 

Among the machine learning methods~\cite{xia20},~\cite{ye2017power}, variational autoencoder (VAE) is a generative model that is widely applied in unsupervised and supervised learning~\cite{zhao2021variational}. VAE is used in many areas, including image processing~,\cite{turhan2018variational}, text data processing~\cite{miao2016neural}, and recommendation systems~\cite{liang2018variational}. A method using VAE  for blind channel equalization has also been proposed in~\cite{caciularu2018blind}, and this study demonstrates the benefits of VAE for solving the precoding (beamforming) problem in the wireless MIMO system. A method using VAE in the field of data transmission is proposed in~\cite{lopez2019variational}. 

This work studies the potential of variational autoencoder (VAE) and conditional variational autoencoder (CVAE) in wireless MIMO systems as deep learning precoding methods. Both VAE and CVAE methods can be used to learn the distribution of precoding matrices with high spectral efficiency. Models of VAE and CVAE are based on the encoder-decoder structure and also utilize latent variables sampled from the standard normal distribution in order to assist decoding and sampling. The framework of variational optimization is constructed. Based on the theory of Bayesian inference, the parameters of the VAE model are obtained through the maximization of Variational Lower Bound (VLB). The proposed VAE/CVAE approach is compared with the complex and highly efficient optimization of L-BFGS~\cite{LBFGS}~\cite{bobrov2021study}. In the case of a single antenna, which is a special case of MIMO, high values of average SE similar to that of L-BFGS can be achieved using precoding matrices sampled from VAE/CVAE with less execution time than the iterative optimization L-BFGS method. We measure the quality of the models using Signal-to-Interference-and-Noise-Ratio (SINR)~\cite{SINR} and spectral efficiency (SE)~\cite{SE}. 

The remainder of this paper is organized as follows.  In Section~\ref{sec:VAE} we describe the variational autoencoder model. In Section~\ref{sec:MIMO} we describe the system model of the studied Massive MIMO network.  In Section~\ref{sec:Problem} we set up the problem that is being solved. In Section~\ref{sec:Approach} we investigate the distribution of precoding matrices corresponding to the same spectral efficiency values. Here we also consider the problem of finding a good precoding matrix using a particular deep learning model~--- so-called conditional variational autoencoder. Section~\ref{sec:Results} contains numerical results and Section~\ref{sec:Conclusion} contains the conclusion.


\section{Variational Autoencoder}\label{sec:VAE}

Given a set of independent and identically distributed samples from true data distribution  $W_i \sim p_d(W)$, $i = 1, \dots, N$, variational autoencoder (VAE) is used to  build a probabilistic model $p_\theta(W)$ of the true data distribution $p_d(W)$.

VAE is a generative model that uses latent variables for the distribution reconstruction and sampling. Using VAE, it is possible to sample new objects from the model distribution $p_\theta(W)$ in two steps:
\begin{enumerate}
    \item  Sample $Z \sim p(Z)$.
    \item  Sample $W \sim p_\theta(W | Z)$.
\end{enumerate}
Here $Z$ is a latent variable from a fixed prior distribution $p(Z)$. The parameters of the distribution $p_\theta(W | Z)$ are obtained using a neural network with weights $\theta$ and $Z$ as an input, namely $p_\theta(W | Z) = \mathcal{N}(W | \mu_\theta(Z), \sigma^2_\theta(Z)I)$. In this case the output of the network is a pair of mean $\mu_\theta(Z)$ and dispersion $\sigma^2_\theta(Z)$ of the reconstructed distribution. This network is called a generator, or a decoder.

To fit the model to data we maximize marginal log-likelihood $\log p_\theta(W)$ of the train set. However, $\log p_\theta(W)$ cannot be optimized straightforwardly, because there is an integral in high-dimensional space inside the logarithm which cannot be computed analytically or numerically estimated with enough accuracy in a reasonable amount of time~\cite{kingma2013auto}. So, in order to perform optimization, we instead maximize the variational lower bound (VLB) on log-likelihood:
\begin{equation}\label{eq:VLB}
    \log p_\theta(W) \geqslant  \mathbb{E}_{Z \sim q_\phi(Z | W)} \log p_\theta(W | Z) - KL(q_\phi(Z | W) || p(Z)) = L(W; \phi, \theta) \to \max\limits_{\phi, \theta}.
\end{equation}

Here $KL(q||p)$ denotes the Kullback-Leibler divergence which qualifies the distance of two distributions $q$ and $p$~\cite{kingma2013auto}. For two normal distributions, KL-divergence has a closed-form expression. Also $q_\phi(Z | W)$ is called a proposal, recognition or variational distribution. It is usually defined as a Gaussian with parameters from a neural network with weights $\phi$ which takes $W$ as an input: $q_\phi(Z | W) = \mathcal{N}(Z | \mu_\phi(W), \sigma^2_\phi(W)I)$. Again, in this case, the output of the network is a pair of mean $\mu_\phi(W)$ and dispersion $\sigma^2_\phi(W)$ of the proposal distribution. This network is called a proposal network, or an encoder. 

It is also possible to make variational autoencoder dependent on some additional input $H$. Prior distribution over $Z$ is now conditioned on $H$, i.e. $p_\psi(Z | H) = \mathcal{N}(Z | \mu_\psi(H), \sigma^2_\psi(H)I)$, where once again mean $\mu_\psi(H)$ and distribution $\sigma^2_\psi(H)$ are the output of the neural network with parameters $\psi$, which takes $H$ as an input. This network is called a prior network and this modification of VAE is called conditional variational autoencoder (CVAE). Note that CVAE uses three neural networks while VAE uses only two.

The sampling process of CVAE is as follows:
\begin{enumerate}
    \item Sample $Z \sim p_\psi(Z | H)$.
    \item Sample $W \sim p_\theta(W | Z, H)$.
\end{enumerate}

Here $Z$ is a latent variable from the prior distribution $p_\psi(Z | H)$, but this distribution is no longer fixed and now depends on the condition $H$ and parameters $\psi$. Proposal and generative networks also have $H$ as an input in CVAE. 

As before, to fit the model to the data we maximize marginal conditional log-likelihood $\log p_{\psi,\theta}(W|H)$ of the train set:
\begin{multline}\label{eq:CVLB}
   \log p_{{\psi},\theta}(W|{H}) \geqslant \mathbb{E}_{Z \sim q_\phi(Z | W, {H})} \log p_\theta(W | Z, {H}) -  KL(q_\phi(Z | W, {H}) || p_{\psi}(Z | {H}))  = \\ = L(W; \phi, {\psi}, \theta)
\to \max\limits_{\phi, {\psi}, \theta}. 
\end{multline}


\section{Optimization Function}\label{sec:MIMO}

\begin{figure}
    \centering
    \includegraphics[width=0.95\linewidth]{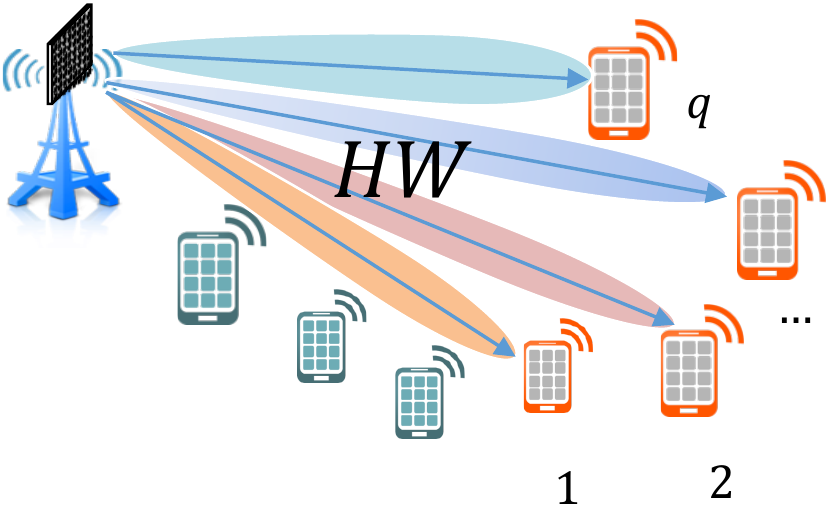}
    \caption{Multi-User precoding allows to transmit different information to different users simultaneously. Using the matrix $W$ we can configure the amplitude and phase of the beams presented on the picture. The problem is to approximate the  precoding matrix $W$ using machine learning as fast as possible, given the SE target function~\eqref{eq:problem1} assuming that the data set $W$ is constructed in advance.}
    \label{fig:mimo_example}
\end{figure}

Consider a  MIMO setting with a single antenna for each user, where the total number of users is $q$, and $k = 1 \dots q$ is a user index. The setting is characterized by the channel matrix $H$; $h_k \in \mathbb{C}^{n}$ is $k$-th row of $H \in \mathbb{C}^{q \times n}$, and $n$ is the number of base station antennas. Given a channel matrix $H$ we need to find precoding matrix~$W$; $w^k \in \mathbb{C}^{n}$ is $k$-th column of $W \in \mathbb{C}^{n \times q}$. 

The quality of transmission is determined by $\mathrm{SINR}_k(W)$~--- signal-to-interference-and-noise ratio (SINR), the ratio of useful signal to interference signal plus noise:
\begin{equation}
    \mathrm{SINR}_k(W) = \frac{|h_k\cdot w^k|^2}{\sum_{l\neq k}|h_k\cdot w^l|^2 + \sigma^2},
\end{equation}
where $\sigma^2$ is the noise power of the system.

The criterion for estimating the quality of precoding is spectral efficiency (SE) $f(W)$~\cite{SE}. We should account for a constraint on the signal power emitted from $i$-th antenna $\sum_k|w^k_i|^2$. The optimization problem for finding precoding $W\in\mathbb{C}^{n{\times}q}$:
\begin{align}\label{eq:problem1}
	&f(W) = \sum_k\log_2\big(1 +  \mathrm{SINR}_k(W)\big)\rightarrow\max_W \\
	&\max_{1\le i\le n}\sum_l|w^l_i|^2 \le p\notag
\end{align}

An approximate solution of~\eqref{eq:problem1} can be obtained using L-BFGS method~\cite{bobrov2021study}.

One can notice that if $W$ is the solution matrix for the above optimization problem then for any diagonal matrix $Q$ which elements lie on the unit circle in the complex plane, the matrix $W \cdot Q$ produces the same spectral efficiency and satisfies per-antenna power constraints. To study the distribution of precoding matrices with high SE we decided to use a variational autoencoder.


\section{Problem Setup and Research Questions}\label{sec:Problem}

To sum up, in this task our goal is to train a variational autoencoder neural network, conditioned on channel matrix $H$, that generates precoding matrices for MIMO problems with channel matrix $H$.
Our research questions are as follows:
\begin{itemize}
    \item Can we produce new high-SE precoding matrices for $H$ by training variational autoencoder on a set of precoding matrices $W_j$, corresponding to a fixed channel matrix $H$?
    \item Can we produce high-SE precoding matrices using variational autoencoder in the case of arbitrary channel matrix $H$, i. e. by conditioning VAE on $H$?
\end{itemize}

\begin{figure}
    \centering
    \includegraphics[width=0.87\linewidth]{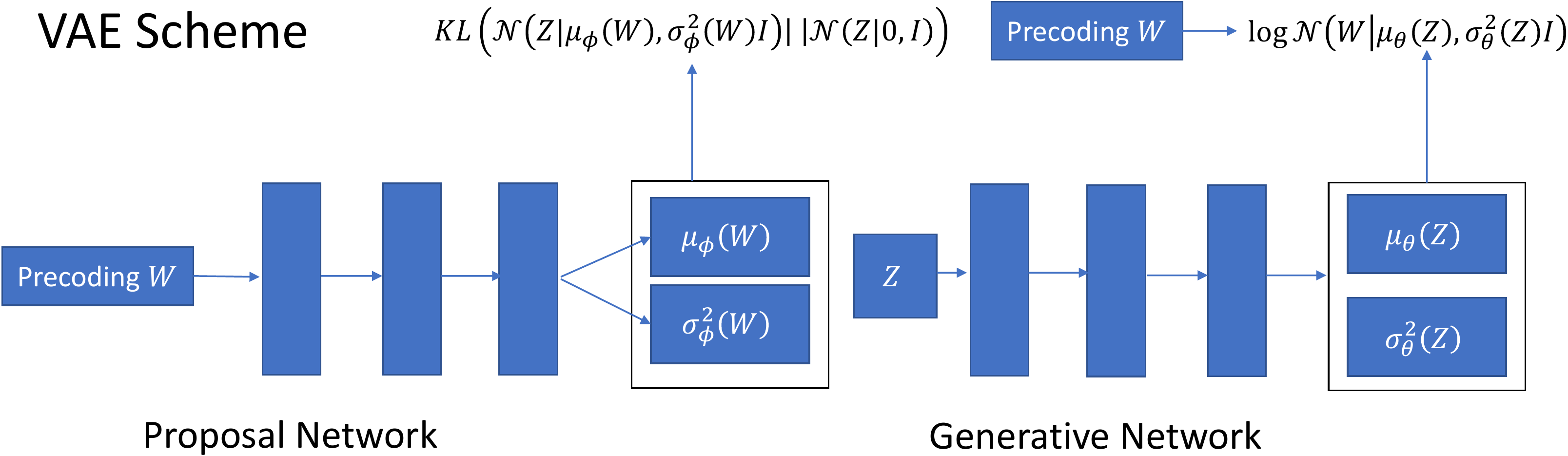}
    \caption{The neural network model of VAE. \textbf{Training}. Encoder (proposal network) takes matrix $W$ as input and produces vector $\mu_\phi(W)$ of shape $d_{model}$ and a single number $\sigma_{\phi}^2(W)$. A latent code $Z$ is sampled from $\mathcal{N}(Z | \mu_\phi(W), \sigma_{\phi}^2(W)I)$ using reparametrization $Z = \mu_\phi(W) + \varepsilon \cdot \sigma_{\phi}(W)$, where $\varepsilon$ is standard gaussian noise. Decoder (generative network) takes previously sampled latent code $Z$ and produces vector $\mu_\theta(Z)$ of shape $d_{data}$ and a single number $\sigma_{\theta}^2(Z)$ which are the parameters of a normal distribution $\mathcal{N}(W | \mu_\theta(Z), \sigma^2_\theta(Z) I)$. Finally, VAE VLB~\eqref{eq:VLB} is computed and optimizer step is made. \textbf{Sampling}. Using VAE model we sample latent code $Z$ from standard complex gaussian noise $\mathcal{N}(Z |0, I)$. We then use this code to obtain decoder output $p_\theta(Z|W)$ and treat it as a sampled precoding matrix $W$. Remember that VAE model can  be trained and produce samples of $W$ only for one specific channel matrix $H$. To produce precoding matrix $W$ for the arbitrary channel $H$ a CVAE model should be used (see Fig.~\ref{fig:nn_model}).}
    \label{fig:vae_nn_model}
\end{figure}


\begin{figure}
    \centering
    \includegraphics[width=0.87\linewidth]{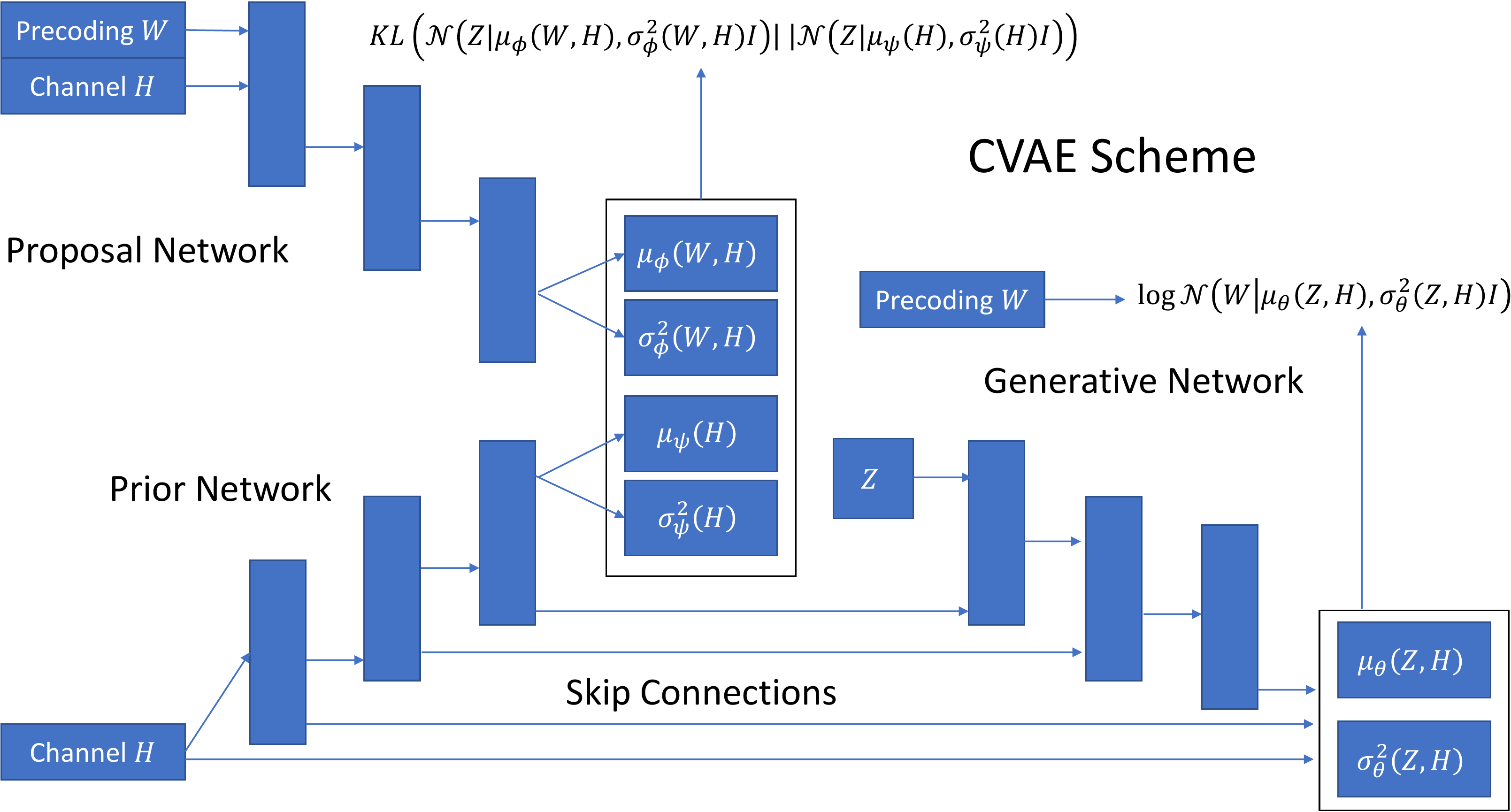}
    \caption{The neural network model of CVAE. \textbf{Training}. Encoder (proposal network) takes matrices $W$ and $H$ as input and produces vector $\mu_\phi(W, H)$ of shape $d_{model}$ and a single number $\sigma_{\phi}^2(W, H)$. Prior network takes matrix $H$ as input and produces vector $\mu_\psi(W, H)$ of shape $d_{model}$ and a single number $\sigma_{\psi}^2(W, H)$. A latent code $Z$ is sampled from $\mathcal{N}(Z | \mu_\phi(W, H), \sigma_{\phi}^2(W, H)I)$, using reparametrization. Decoder (generative network) takes previously sampled latent code $Z$, matrix $H$  and produces vector $\mu_\theta(Z, H)$ of shape $d_{data}$ and a single number $\sigma_{\theta}^2(Z, H)$ which are the parameters of a normal distribution $\mathcal{N}(W | \mu_\theta(Z, H), \sigma^2_\theta(Z, H) I)$. Finally, CVAE VLB~\eqref{eq:CVLB} is computed and optimizer step is made. Also skip connections solve the degradation problem in neural network training~\cite{he2016deep}. \textbf{Sampling}. Given matrix $H$ we sample latent code $Z$ from prior distribution $p_\psi(Z | H)$ parametrized with prior network. We then use this code to obtain decoder output $p_\theta(W | Z, H)$ and treat it  as a sampled precoding $W$ for channel $H$.}
    \label{fig:nn_model}
\end{figure}

\section{Approach}\label{sec:Approach}

In this section, we describe the training procedure for (conditional) variational autoencoder.

We begin with the unconditional one. 
Variational autoencoder consists of two networks: encoder $q_\phi(Z | W)$ and decoder $p_\theta(W | Z)$. 
During training, for each matrix $W$:
\begin{enumerate}
    \item Encoder takes matrix $W$ as input and produces vector $\mu_\phi(W)$ of shape $d_{model}$ and a single number $\sigma_{\phi}^2(W)$. 
    \item A latent code $Z$ is sampled from $\mathcal{N}(Z | \mu_\phi(W), \sigma_{\phi}^2(W)I)$ using reparametrization $Z = \mu_\phi(W) + \varepsilon \cdot \sigma_{\phi}(W)$, where $\varepsilon$ is standard gaussian noise.
    \item Decoder takes previously sampled latent code $Z$ and produces   vector $\mu_\theta(Z)$ of shape $d_{data}$ and a single number $\sigma_{\theta}^2(Z)$ which are the parameters of a normal distribution $\mathcal{N}(W | \mu_\theta(Z), \sigma^2_\theta(Z) I)$.  
    \item Compute VLB~\eqref{eq:VLB} and make optimizer step.
\end{enumerate}
We fix $d_{model}$~--- the dimension of latent space~--- to be equal to $64$ and prior distribution $p(Z)$ to be standard gaussian distribution in $d_{model}$-dimensional space.
We also represent complex matrix $W$ of shape $16 \times 4$ as a real vector of shape $d_{data} = 16 \times 4 \times 2 = 128$ of real and imaginary parts of $W$.

To generate possible matrices $W$ after training, we simply sample $Z$ from prior distribution $p(Z)$ and then pass it through the decoder network $p_\theta(W|Z)$, treating $\mu_\theta(Z)$ as a sampled matrix $W$.

Having trained unconditional autoencoder, it is almost straightforward to obtain a conditional one.
In addition to encoder and decoder networks $p_\theta(W| Z, H)$ and $q_\phi(Z| W, H)$, CVAE also contains a prior network $p_\psi(Z | H)$. As previously, $d_{data}$ is equal to $128 = 16 \times 4 \times 2$ and $d_{model} = 64$.
To make a training step, we do the following:
\begin{enumerate}
    \item Encoder (proposal network) takes matrices $W$ and $H$ (represented as two real-valued vectors of shape $128$ each) as input and produces vector $\mu_\phi(W, H)$ of shape $d_{model}$ and a single number $\sigma_{\phi}^2(W, H)$.
    \item Prior network takes matrix $H$ as input and produces vector $\mu_\psi(H)$ of shape $d_{model}$ and a single number $\sigma_{\psi}^2(H)$.
    \item A latent code $Z$ is sampled from $\mathcal{N}(Z | \mu_\phi(W, H), \sigma_{\phi}^2(W, H)I)$ using repara-metrization.
    \item Decoder (generative network) takes previously sampled latent code $Z$, matrix $H$ (represented as real-valued vector of shape $128$) and produces vector $\mu_\theta(Z, H)$ of shape $d_{data}$ and a single number $\sigma_{\theta}^2(Z, H)$ which are the parameters of a normal distribution $\mathcal{N}(W | \mu_\theta(Z, H), \sigma^2_\theta(Z, H) I)$.
    \item Computing VLB for CVAE~\eqref{eq:CVLB} and making optimizer step.
\end{enumerate}
Usage of CVAE for sampling matrices $W$, however, is slightly more difficult. 
Given matrix $H$ we sample latent code $Z$ from prior distribution $p_\psi(Z | H)$ parametrized with prior network.
We then use this code to obtain decoder output $\mu_\theta(W | Z, H)$ and treat it  as a sampled precoding matrix $W$ for channel $H$.
The neural network model of CVAE is presented in Fig.~\ref{fig:nn_model}. 

\section{Experimental Results}\label{sec:Results}

In this section, we present the results of training VAE and CVAE to produce precoding matrices with high spectral efficiency for a given channel matrix $H$. First of all, we report the results of training VAE to fit a distribution of precoding matrices $W$ for a fixed channel $H$. We consider two approaches: VAE and a CVAE with the fixed matrix $H$ as a condition.

To obtain target precoding matrices $W$ for a given $H$, we optimize spectral efficiency using the L-BFGS optimization approach~\cite{bobrov2021study}, starting from randomly initialized matrix $W$, until convergence. The L-BFGS method is implemented with the PyTorch~\cite{PyTorch} framework.

\begin{figure}
    \centering
    \includegraphics[width=1\linewidth]{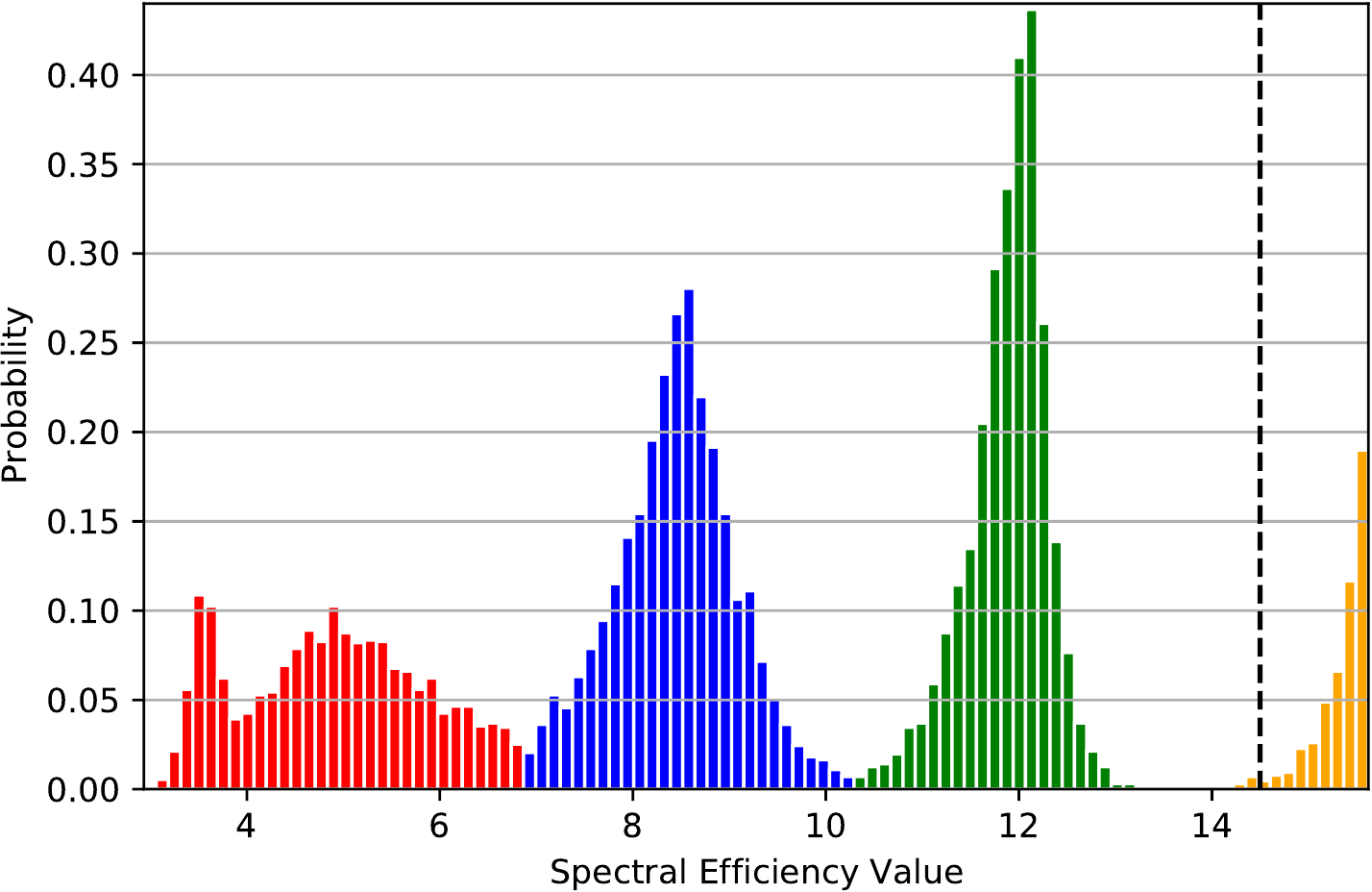}
    \caption{Distribution of spectral efficiency for precoding matrices $W_i$ obtained with random initialization of L-BFGS method for a fixed channel matrix $H$. Orange corresponds to samples with high efficiency (14 SE and more), blue and green~--- samples with average efficiency  (from 7 to 14 SE), red~--- samples with low efficiency (less than 7 SE).  \textbf{Training}. Samples with SE above threshold $t = 14.5$ (dashed line) were selected for training VAE/CVAE neural networks. Using such training data and following the experiment shown in Sec.~\ref{sec:fixed_H} and Fig.~\ref{fig:SE}, it is possible to generate highly efficient precoding matrices $W$.}
    \label{fig:quality}
\end{figure}

\begin{figure}
    \centering
    \includegraphics[width=1\linewidth]{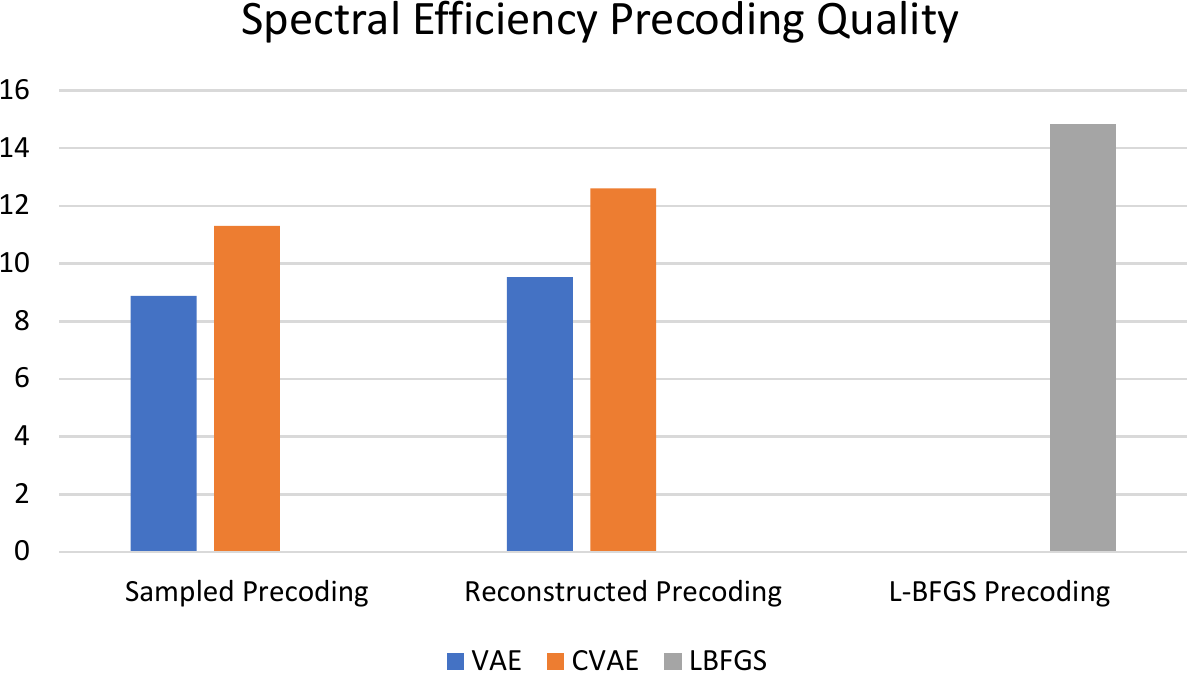}
    \caption{Spectral efficiency precoding quality produced by the algorithms.}
    \label{fig:SE}
\end{figure}

\subsection{Generating high SE precoding matrices $W_i$ for the fixed $H$}\label{sec:fixed_H}

For a given channel matrix $H \in \mathbb{C}^{q{\times}n}$, we find the precoding $W \in \mathbb{C}^{n \times q}$ using L-BFGS approach~\cite{bobrov2021study} with a random initialization. This method returns precoding matrices with high spectral efficiency using the iterative optimization approach and measures the SE quality using the formula of SE~\eqref{eq:problem1}. We repeat this from different starting points to get different matrices $W_j$ for a fixed $H$, filtering out precoding matrices with SE lower than a certain threshold $t$ in the process (see Fig.~\ref{fig:quality}), obtaining a set $\{W_j\}_{j=1}^{N}$ of size $N$.

We measured the quality of generated precoding matrices $W \in \mathbb{C}^{n \times q}$ simply by evaluating spectral efficiency~\eqref{eq:problem1} $f(W)$  with a given channel $H  \in \mathbb{C}^{q{\times}n}$. For a given channel matrix $H$ and a set of $N=5000$ precoding matrices $W_i$, the average SE value for $W_i$ and $H$ is $14.85$ (Fig.~\ref{fig:SE} L-BFGS Precoding). We fix matrix dimensions to be equal to $q=4, n=16$, and the value of filtering threshold $t=14.5$.

We managed to obtain the mean SE value of $8.87$ with VAE and $11.31$ with CVAE for a fixed $H$ matrix for sampled precoding matrices $W_i$ (Fig.~\ref{fig:SE} Sampled Precoding). This method samples from the distribution of precoding matrices with high SE~\eqref{eq:problem1} using the deep learning approach of VAE/CVAE.  We also measured mean SE in reconstructed objects (precodings $W_i$ passed through encoder and decoder) and obtained $9.53$ and $12.61$ for VAE and CVAE-fixed-$H$ respectively (Fig.~\ref{fig:SE} Reconstructed Precoding).

\subsection{Generating matrices $W_{i,j}$ using $H_i$ as condition for CVAE}

To study how MIMO solutions change with respect to $H$, we perturb $H$ with random matrices $\Delta_i$, obtaining $K$ different channel matrices $H_i = H + \Delta_i$, where each $\Delta_i$ satisfies $||\Delta_i||_F^2 = \delta$ and repeat the exact same process of generating a set of $N$ precoding matrices for each $H_i$. Overall, we have a set of pairs $\{H_i, W_{i, j}\}_{i,j=1}^{K + 1, N}$ of channel matrices and corresponding precodings with high spectral efficiency, where $H \in \mathbb{C}^{q{\times}n}$ and $W \in \mathbb{C}^{n \times q}$. We fix hyper-parameters to be equal to $K=15, q=4, n=16, N=500, \delta=5$. For this task, channel matrices $H$ are generated from the complex standard normal distribution.

We train CVAE on a set $\{H_i, W_{i,j}\}$ with $H_i = H + \Delta_i$ and $H_0 = H$ for convenience. However, we do not manage to get high-quality samples from such a model, having mean SE for reconstructed objects equal to $6.49$ and $6.45$ if we use one of $H_i$ as a condition and we observe the same fall in spectral efficiency for a perturbation of $H$ which differ from one of $\Delta_i$.



\subsection{Computational Complexity}

The proposed VAE/CVAE solution works much faster than the L-BFGS reference method. We estimate the inference of the proposed two-layers neural network as one iteration of the L-BFGS procedure. Which includes two matrix-vector operations inside the layers and one matrix-vector multiplication of SE~\eqref{eq:problem1}. The complexity of the L-BFGS iteration is estimated as three SE~\eqref{fig:SE} estimations~\cite{SE}. Since in practice the number of iterations reaches up to a hundred~\cite{LBFGS}, the inference of the proposed neural network is faster up to a hundred times. Both VAE and CVAE methods require a training procedure in the background, but we mainly focus on the inference time in a real system. The learning time of the CVAE method is higher than that of the VAE method~\cite{pagnoni2018conditional}.  If we denote the complexity of the VAE/CVAE inference as $y$, the relative complexity of the L-BFGS method will be $My$, where $M$ is the number of L-BFGS iterations.

\section{Conclusion}\label{sec:Conclusion}
In this paper, we propose a deep learning algorithm based on the theory of variational autoencoders. This algorithm is designed to generate precoding matrices with high spectral efficiency for massive MIMO systems. With its help, we successfully sampled precoding matrices using both versions of the autoencoder: VAE and CVAE. The first one must be trained separately for different channel matrices $H$, and the second one can be conditioned on the matrix of the channel $H$ and thus can be used for different $H$ as a unified model. In the case of a fixed channel matrix $H$, both approaches were able to generate precoding matrices with high spectral efficiency, but CVAE surpasses VAE in terms of quality and versatility. However, both approaches can be used to study the distribution of precoding matrices with high spectral efficiency.

\section*{Acknowledgements}
The authors are grateful to Mr.~D.~Kropotov and Prof.~O.~Senko for discussions.

%
%
%
\bibliographystyle{splncs04}
\bibliography{mybibliography}

\begin{thebibliography}{10}
\providecommand{\url}[1]{\texttt{#1}}
\providecommand{\urlprefix}{URL }
\providecommand{\doi}[1]{https://doi.org/#1}

\bibitem{abdallah2018mmse}
Abdallah, A., Mansour, M.M., Chehab, A., Jalloul, L.M.: {MMSE} detection for
  1-bit quantized massive {MIMO} with imperfect channel estimation. In: 2018
  IEEE 19th International Workshop on Signal Processing Advances in Wireless
  Communications (SPAWC). pp.~1--5. IEEE (2018)

\bibitem{albreem2019massive}
Albreem, M.A., Juntti, M., Shahabuddin, S.: Massive {MIMO} detection
  techniques: A survey. IEEE Communications Surveys \& Tutorials
  \textbf{21}(4),  3109--3132 (2019)

\bibitem{5G}
Andrews, J.G., Buzzi, S., Choi, W., Hanly, S.V., Lozano, A., Soong, A.C.,
  Zhang, J.C.: What will {5G} be? IEEE Journal on selected areas in
  communications  \textbf{32}(6),  1065--1082 (2014)

\bibitem{balatsoukas2019neural}
Balatsoukas-Stimming, A., Casta{\~n}eda, O., Jacobsson, S., Durisi, G., Studer,
  C.: Neural-network optimized 1-bit precoding for massive {MU-MIMO}. In: 2019
  IEEE 20th International Workshop on Signal Processing Advances in Wireless
  Communications (SPAWC). pp.~1--5. IEEE (2019)

\bibitem{bjornson2016massive}
Bj{\"o}rnson, E., Larsson, E.G., Marzetta, T.L.: {Massive} {MIMO}: {Ten} myths
  and one critical question. IEEE Communications Magazine  \textbf{54}(2),
  114--123 (2016)

\bibitem{MCS}
Bobrov, E., Kropotov, D., Lu, H., Zaev, D.: Massive mimo adaptive modulation
  and coding using online deep learning algorithm. IEEE Communications Letters
  \textbf{26}(4),  818--822 (2022). \doi{10.1109/LCOMM.2021.3132947}

\bibitem{bobrov2021study}
Bobrov, E., Kropotov, D., Troshin, S., Zaev, D.: {Study on Precoding
  Optimization Algorithms in Massive {MIMO} System with Multi-Antenna Users}
  (2021)

\bibitem{bobrov2021machine}
Bobrov, E., Troshin, S., Chirkova, N., Lobacheva, E., Panchenko, S., Vetrov,
  D., Kropotov, D.: Machine learning methods for spectral efficiency prediction
  in massive mimo systems (2021)

\bibitem{caciularu2018blind}
Caciularu, A., Burshtein, D.: Blind channel equalization using variational
  autoencoders. In: 2018 IEEE International Conference on Communications
  Workshops (ICC Workshops). pp.~1--6. IEEE (2018)

\bibitem{dong2020framework}
Dong, P., Zhang, H., Li, G.Y.: Framework on deep learning-based joint hybrid
  processing for {mmWave} massive {MIMO} systems. IEEE Access  \textbf{8},
  106023--106035 (2020)

\bibitem{he2016deep}
He, K., Zhang, X., Ren, S., Sun, J.: Deep residual learning for image
  recognition. In: Proceedings of the IEEE conference on computer vision and
  pattern recognition. pp. 770--778 (2016)

\bibitem{huang2020deep}
Huang, C., Chen, G., Gong, Y., Xu, P.: Deep reinforcement learning based relay
  selection in delay-constrained secure buffer-aided {CRNs}. In: GLOBECOM
  2020-2020 IEEE Global Communications Conference. pp.~1--6. IEEE (2020)

\bibitem{kingma2013auto}
Kingma, D.P., Welling, M.: Auto-encoding variational bayes. arXiv preprint
  arXiv:1312.6114  (2013)

\bibitem{larsson2014massive}
Larsson, E.G., Edfors, O., Tufvesson, F., Marzetta, T.L.: Massive {MIMO} for
  next generation wireless systems. IEEE communications magazine
  \textbf{52}(2),  186--195 (2014)

\bibitem{liang2018variational}
Liang, D., Krishnan, R.G., Hoffman, M.D., Jebara, T.: Variational autoencoders
  for collaborative filtering. In: Proceedings of the 2018 world wide web
  conference. pp. 689--698 (2018)

\bibitem{lopez2019variational}
Lopez-Martin, M., Carro, B., Sanchez-Esguevillas, A.: Variational data
  generative model for intrusion detection. Knowledge and Information Systems
  \textbf{60}(1),  569--590 (2019)

\bibitem{miao2016neural}
Miao, Y., Yu, L., Blunsom, P.: Neural variational inference for text
  processing. In: International conference on machine learning. pp. 1727--1736.
  PMLR (2016)

\bibitem{EE}
Ngo, H.Q., Larsson, E.G., Marzetta, T.L.: Energy and spectral efficiency of
  very large multiuser {MIMO} systems. IEEE Transactions on Communications
  \textbf{61}(4),  1436--1449 (2013)

\bibitem{pagnoni2018conditional}
Pagnoni, A., Liu, K., Li, S.: Conditional variational autoencoder for neural
  machine translation. arXiv preprint arXiv:1812.04405  (2018)

\bibitem{ZF_MRT}
Parfait, T., Kuang, Y., Jerry, K.: Performance analysis and comparison of {ZF}
  and {MRT} based downlink massive mimo systems. In: 2014 sixth international
  conference on ubiquitous and future networks (ICUFN). pp. 383--388. IEEE
  (2014)

\bibitem{PyTorch}
Paszke, A., Gross, S., Chintala, S., Chanan, G., Yang, E., DeVito, Z., Lin, Z.,
  Desmaison, A., Antiga, L., Lerer, A.: Automatic differentiation in pytorch
  (2017)

\bibitem{rezaie2020location}
Rezaie, S., Manch{\'o}n, C.N., De~Carvalho, E.: Location-and orientation-aided
  millimeter wave beam selection using deep learning. In: ICC 2020-2020 IEEE
  International Conference on Communications (ICC). pp.~1--6. IEEE (2020)

\bibitem{rusek2012scaling}
Rusek, F., Persson, D., Lau, B.K., Larsson, E.G., Marzetta, T.L., Edfors, O.,
  Tufvesson, F.: Scaling up {MIMO}: Opportunities and challenges with very
  large arrays. IEEE signal processing magazine  \textbf{30}(1),  40--60 (2012)

\bibitem{DPC}
Tran, L.N., Juntti, M., Bengtsson, M., Ottersten, B.: Beamformer designs for
  {MISO} broadcast channels with zero-forcing dirty paper coding. IEEE
  transactions on wireless communications  \textbf{12}(3),  1173--1185 (2013)

\bibitem{turhan2018variational}
Turhan, C.G., Bilge, H.S.: Variational autoencoded compositional pattern
  generative adversarial network for handwritten super resolution image
  generation. In: 2018 3rd International Conference on Computer Science and
  Engineering (UBMK). pp. 564--568. IEEE (2018)

\bibitem{van2020deep}
Van~Luong, T., Ko, Y., Vien, N.A., Matthaiou, M., Ngo, H.Q.: Deep energy
  autoencoder for noncoherent multicarrier {MU-SIMO} systems. IEEE Transactions
  on Wireless Communications  \textbf{19}(6),  3952--3962 (2020)

\bibitem{SE}
Verd{\'u}, S.: Spectral efficiency in the wideband regime. IEEE Transactions on
  Information Theory  \textbf{48}(6),  1319--1343 (2002)

\bibitem{SINR}
Wang, B., Chang, Y., Yang, D.: On the {SINR} in massive {MIMO} networks with
  {MMSE} receivers. IEEE Communications Letters  \textbf{18}(11),  1979--1982
  (2014)

\bibitem{xia20}
Xia, W., Zheng, G., Zhu, Y., Zhang, J., Wang, J., Petropulu, A.P.: A deep
  learning framework for optimization of {MISO} downlink beamforming. IEEE
  Transactions on Communications  \textbf{68}(3),  1866--1880 (2019)

\bibitem{ye2017power}
Ye, H., Li, G.Y., Juang, B.H.: Power of deep learning for channel estimation
  and signal detection in {OFDM} systems. IEEE Wireless Communications Letters
  \textbf{7}(1),  114--117 (2017)

\bibitem{ye2019deep}
Ye, H., Li, G.Y., Juang, B.H.F.: Deep reinforcement learning based resource
  allocation for {V2V} communications. IEEE Transactions on Vehicular
  Technology  \textbf{68}(4),  3163--3173 (2019)

\bibitem{zhao2021variational}
Zhao, T., Li, F.: Variational-autoencoder signal detection for {MIMO-OFDM-IM}.
  Digital Signal Processing  \textbf{118},  103230 (2021)

\bibitem{zhao2020deep}
Zhao, T., Li, F., Tian, P.: A deep-learning method for device activity
  detection in {mMTC} under imperfect {CSI} based on variational-autoencoder.
  IEEE Transactions on Vehicular Technology  \textbf{69}(7),  7981--7986 (2020)

\bibitem{LBFGS}
Zhu, C., Byrd, R.H., Lu, P., Nocedal, J.: Algorithm 778: {L-BFGS-B}: {Fortran}
  subroutines for large-scale bound-constrained optimization. ACM Transactions
  on mathematical software (TOMS)  \textbf{23}(4),  550--560 (1997)

\end{thebibliography}

\end{document}